\documentclass[prb,twocolumn,showpacs]{revtex4}
\usepackage{graphicx}
\usepackage{amsmath}
\usepackage{amssymb}

\newcommand{\be}{\begin{equation}}
\newcommand{\ee}{\end{equation}}
\newcommand{\ba}{\begin{eqnarray}}
\newcommand{\ea}{\end{eqnarray}}

\begin{document}
%\preprint{HEP/123-qed}

\title{Unconventional Integer Quantum Hall effect in graphene}

\author{V.P.~Gusynin$^{1}$}
%\thanks{On leave of absence from }
\email{vgusynin@bitp.kiev.ua}
\author{S.G.~Sharapov$^{2}$}
\email{sharapov@bitp.kiev.ua}
%\thanks{Present address: Bogolyubov Institute for Theoretical Physics, Kiev, Ukraine}
%\homepage{}

\affiliation{$^1$ Bogolyubov Institute for Theoretical Physics,
        Metrologicheskaya Str. 14-b, Kiev, 03143, Ukraine\\
        $^2$ Department of Physics and Astronomy, McMaster University,
        Hamilton, Ontario, Canada, L8S 4M1}

\date{\today }

\begin{abstract}
Monolayer graphite films, or graphene,  have  quasiparticle
excitations that can be described by $2+1$ dimensional Dirac theory.
We demonstrate that this produces an unconventional form of the
quantized Hall conductivity $\sigma_{xy} = - (2 e^2/h)(2n+1)$ with
$n=0,1,\ldots$, that notably distinguishes graphene from other
materials where the integer quantum Hall effect was observed. This
unconventional quantization is caused by the quantum anomaly of the
$n=0$ Landau level and was discovered in recent experiments on
ultrathin graphite films.
\end{abstract}


%73.43.-f   Quantum Hall effects
%73.43.Cd   Theory and modeling
%71.70.Di       Landau levels
%81.05.Uw   Carbon, diamond, graphite

%\keywords{QUANTUM HALL EFFECT; GRAPHENE}

\maketitle

The quantum Hall effect (QHE) is one of the most remarkable
phenomena in condensed matter discovered in the second half of the
20th century. The basic experimental fact characterizing QHE is that
the diagonal electric conductivity of a two-dimensional electron
system in a strong magnetic field is vanishingly small
$\sigma_{xx}\to 0$, while the non-diagonal conductivity is quantized
in multiples of $e^2/h$: $\sigma_{xy}=-\nu e^2/h$, where $\nu$ is an
integer (the integer quantum Hall effect (IQHE)) or a fractional
number (the fractional QHE). In a recent paper
\cite{Novoselov2004Science} the fabrication of free-standing
monocrystalline graphite films with thickness down to a single
atomic layer was reported. This new material, called graphene,
possesses truly remarkable properties such as excellent mechanical
characteristics, scalability to the nanometer sizes, and the ability
to sustain huge ($>10^8 A/cm^2$) electric currents. By using the
electric field effect \cite{Novoselov2004Science}, it is possible to
change the carrier concentration in samples by tens times and even
to change the carrier type from electron to hole when the sign of
applied gate voltage is altered. All this make graphene a promising
candidate for applications in future micro- and nanoelectronics.

On the theoretical side, the linear, Dirac-like, spectrum of
quasiparticle excitations (up to energies of the order of $1000$\,K)
and the pseudospin degeneracy make graphene a unique truly
two-dimensional "relativistic" electronic system. The thinnest
graphite films can be described by a low-energy (2+1) dimensional
effective {\em massless\/} Dirac theory \cite{Semenoff1984PRL}. Of
special interest are the properties of graphene in a magnetic field.
The important differences between the Dirac and Schr\"odinger
theories may be observed in thermodynamic and magnetotransport
measurements
\cite{Kopelevich2003PRL,Novoselov2004Science,Morozov2005,Berger2004JPCB,Zhang2005PRL}.
For instance, the phase of de Haas van Alphen and Shubnikov de Haas
oscillations for Dirac quasiparticles is shifted
\cite{Sharapov2004PRB,Gusynin2005PRB,Luk'yanchuk2004PRL,Geim2005}
compared to the phase of non-relativistic quasiparticles. Moreover,
the Dingle and temperature factors in the amplitude of oscillations
explicitly depend on the carrier density in the case of a Dirac-like
spectrum \cite{Sharapov2004PRB,Gusynin2005PRB}.

Because of the large value of the cyclotron gap, it is expected that
the QHE in this material can be observed for much higher
temperatures and lower magnetic fields than in conventional
semiconductors. Therefore it is naturally to ask whether  the fundamental
difference between the properties of Landau levels (LL)
(see Eqs.~(\ref{Dirac-LL}) and (\ref{nonrelLL}) below)
in the Dirac and
Schr\"odinger theories can be observed experimentally in the Hall
conductivity? The purpose of this letter is to show that the
Dirac-like dynamics of graphene results in an unconventional
form of the Hall quantization
\begin{equation}
\label{Hall-Dirac} \sigma_{xy} = - \frac{2 e^2}{h}(2n+1), \quad n =
0,1,\ldots
\end{equation}
We argue that the quantization rule (\ref{Hall-Dirac}) is caused
by the {\em quantum anomaly of the $n=0$ LL\/} , i.e. by the fact
that it has {\em a twice smaller degeneracy\/} than the levels
with $n>0$ and its energy does not depend on the magnetic field
\cite{Gusynin1995PRD}. Remarkably this quantization {\em is
observed\/} experimentally \cite{Geim2005} for ultrathin graphite
films which exhibit the behavior expected for ideal 2D graphene.

We begin with the Lagrangian density of noninteracting
quasiparticles in a single graphene sheet that in the continuum
limit reads \cite{Semenoff1984PRL}
\begin{equation}
\label{Lagrangian} \mathcal{L} \! = \! \sum_{\sigma= \pm 1}
\bar{\Psi}_{\sigma} \left[ i \gamma^0 (\hbar \partial_t - i
\mu_\sigma) + i v_F \gamma^i (\hbar \partial_i - i \frac{e}{ c}A_i)
\right] \Psi_{\sigma},
\end{equation}
where $\Psi_{\sigma} = (\psi_{1\sigma}(t, \mathbf{r}), \psi_{2
\sigma}(t, \mathbf{r}))$ is the four-component Dirac spinor combined
from two spinors $\psi_{1\sigma}, \psi_{2 \sigma}$ [corresponding to
$\mathbf{K}$ and $\mathbf{K}^\prime$ points of the Fermi surface,
respectively]  that describe the Bloch states residing on the two
different sublattices of the biparticle hexagonal lattice of the
graphene sheet, and $\sigma = \pm 1$ is the spin. In
Eq.~(\ref{Lagrangian}) $\gamma^\mu$ with $\mu=0,1,2$ are $4\times 4$
$\gamma$ matrices belonging to a reducible representation in $2+1$,
$\bar{\Psi}_\sigma = \Psi_{\sigma}^\dagger \gamma_0$ is the Dirac
conjugated spinor, $-e<0$ is the  electron charge, $v_F$ is the
Fermi velocity. We set $k_B = 1$, but kept Planck constant $\hbar =
h/2\pi$.

The external magnetic field $\mathbf{B}$ is applied perpendicularly
to the plane along the positive z axis and the vector potential is
taken in the symmetric gauge $\mathbf{A} = (- B/2 y, B/2 x )$. In
contrast to the truly relativistic $(3+1)$ case \cite{foot1}, the
Zeeman interaction term still has to be explicitly added to the
Lagrangian (\ref{Lagrangian}), because it originates from
nonrelativistic many-body theory. This can be done by considering
spin splitting $\mu_{\sigma} = \mu - \sigma \mu_B B$ of the chemical
potential $\mu$, where $\mu_B=e \hbar/(2mc)$ is the Bohr magneton.
However, for realistic values of $v_F \varpropto 10^5 \mbox{m/s}$ in
graphene the distance between LL is very large compared to the
Zeeman splitting \cite{Gusynin2005PRB}, so that in what follows we
will not consider this term and simply multiply all relevant
expressions by 2 to count the spin degeneracy. While simple
tight-binding calculations  made for the hexagonal lattice of a
single graphene sheet predict that $\mu =0$, the real picture is
more complicated and the actual value of $\mu$ can be nonzero due to
finite doping and/or disorder. Moreover, nonzero and even tunable
value of $\mu$ [including the change of the character of carriers,
either electrons or holes] is possible in electric-field doping
experiments \cite{Novoselov2004Science,Morozov2005,Geim2005}. In our
notations $\mu>0$ corresponds to electrons  and accordingly to the
positive gate voltage.

\begin{figure}[h]
\centering{
\includegraphics[width=7cm]{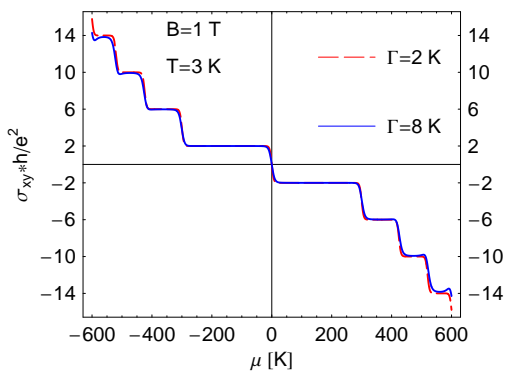}}
\caption{(Color online) The Hall conductivity $\sigma_{xy}$
measured in $e^2/h$ units as a function of chemical potential
$\mu$ for two different values of $\Gamma$ for $T = 3\mbox{K}$ and
$B = 1 \mbox{T}$. We use $\hbar v_F^2eB/c \to (4.5\times10^4 \mbox{K}^2)
B(\mbox{T})$.} \label{fig:1}
\end{figure}

Using the Kubo formalism and modeling the LL by Lorentzians with a
constant width $\Gamma$ the following expression for the diagonal
conductivity was obtained in
Refs.~\cite{Gorbar2002PRB,Gusynin2005PRB}
\begin{equation}
\label{sigma_xx}
\sigma_{xx}(B,\mu,\Gamma) = \frac{2e^2}{\hbar}
\int_{-\infty}^{\infty} d \omega [-n_F^{\prime}(\omega-\mu)]
\mathcal{A}_{xx}(\omega,B,\Gamma),
\end{equation}
where $n_F(\omega)=1/[\exp(\omega/T)+1]$ is the Fermi distribution
and the function $\mathcal{A}_{xx}$ that incorporates the effect of
all LL is given by Eq.~(11) of Ref.~\cite{Gusynin2005PRB}. Now this
result is extended for the Hall conductivity and we derive a general
analytical expression for $\sigma_{xy}(B,\mu,\Gamma)$.
\cite{Gusynin2005} The resulting dependence $\sigma_{xy}(\mu)$ is
shown in Fig.~\ref{fig:1}, where one  sees that the plateaux of
$\sigma_{xy}$ follow Eq.~(\ref{Hall-Dirac}). This agrees with the
latest experimental results \cite{Geim2005} and resemble earlier
theoretical predictions \cite{Zheng2002PRB}.
However, to demonstrate result (\ref{Hall-Dirac}) in the most
transparent way it is useful to write down a simpler conventional
representation \cite{Schakel1991PRD,Gorbar2002PRB} for $\sigma_{xy}$
obtained in the clean limit $\Gamma \to 0$:
\begin{equation}
\label{hall-clean} \sigma_{xy}=-\frac{ec\rho}{B} \equiv -\frac{e^2
\mbox{sgn}(eB) \mbox{sgn} \mu}{\pi \hbar} \nu_B.
\end{equation}
Here we introduced the filling factor of LL, $\nu_B = \pi \hbar
c|\rho| /|eB|$ with $\rho$ being the carrier imbalance
($\rho\equiv n_e - n_h$, where $n_e$ and $n_h$ are the densities
of ``electrons'' and ``holes'', respectively). This filling factor
can be represented as a sum over LL
\begin{equation}
\label{Dirac-LL} M_{n}=\sqrt{\Delta^{2}+ 2 n \hbar v_F^2|eB|/c},\,
\quad n=0,1,\ldots
\end{equation}
of the Dirac theory:
\begin{equation}
\label{filling}
\begin{split}
\mbox{sgn} \mu \, \nu_B & = \frac{1}{2}\left[\tanh
\frac{\mu+\Delta}{2T}+\tanh\frac{\mu-\Delta}{2T} \right.\\
& \left. +2\sum\limits_{n=1}^\infty
\left(\tanh\frac{\mu+M_n}{2T}+\tanh\frac{\mu-M_n}{2T}\right)\right],
\end{split}
\end{equation}
where we separated out the level with $n=0$ because its degeneracy
is only half of  the degeneracy of the levels with $n>0$. To
illustrate this rather peculiar property of the Dirac theory in a
perspicuous way, we included in $M_n$ and $\nu_B$ the mass
(excitonic gap) $\Delta$ which was discussed recently to explain
some experiments \cite{Khveshchenko2001PRL,Gorbar2002PRB}. Our
consideration of $\sigma_{xy}$ is in fact independent of the
presence of $\Delta$, so in what follows we set $\Delta=0$. A zero
value of $\Delta$ is expected for noninteracting quasiparticles on
the hexagonal lattice of graphene.

\begin{figure}[h]
\centering{
\includegraphics[width=7cm]{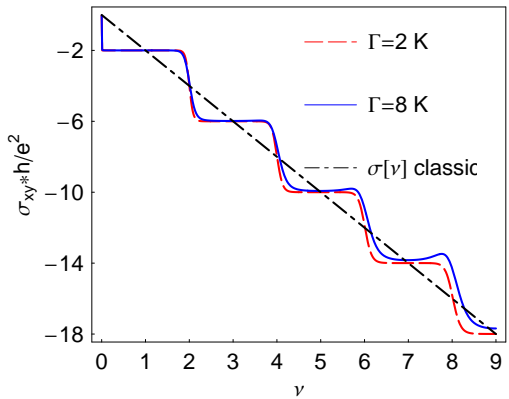}}
\caption{(Color online) The Hall conductivity $\sigma_{xy}$
measured in $e^2/h$ units as a function of the filling $\nu_B$.
The straight line corresponds to a classical dependence
(\ref{hall-clean}). The parameters are the same as in
Fig.~\ref{fig:1}.} \label{fig:2}
\end{figure}

The first equality in Eq.~(\ref{hall-clean}) corresponds to a
classical straight line $\sigma_{12}\varpropto \nu_B$. As discussed,
for example, in Ref.~\cite{Hajdu.book}, this line emerges from two
step function dependences, viz. $\mu(n)$ and $\sigma_{12}(\mu)$.
Indeed, using $\tanh(\omega/2T)={\rm sgn}(\omega)$ for $T \to 0$, we
obtain from Eqs.~(\ref{hall-clean}) and (\ref{filling}) (compare
with Ref.~\cite{Schakel1991PRD})
\begin{equation}\label{sigma-steps}
\sigma_{xy} =
%-\frac{2 e^2}{h} \left[1+ 2 \sum_{n=1}^{\infty}\theta(\mu -M_n)\right]=
-\frac{2e^2\mbox{sgn}(eB)
\mbox{sgn}\mu}{h}\left(1+2\left[\frac{\mu^2c} {2\hbar
|eB|v_F^2}\right]\right),
\end{equation}
where $[x]$ denotes the integer part of $x$. The usual
argumentation (see e.g. Ref.~\cite{Hajdu.book}) for  the
occurrence of the IQHE states that in the presence of disorder the
dependence of $\sigma_{12}(\mu)$ remains the same, while $\mu(n)$
becomes a smooth function. The classical (\ref{hall-clean}) and
quantum (\ref{sigma-steps}) Hall conductivities coincide only for
the fillings, $\nu_B = 2n+1$ (see Fig.~\ref{fig:2}).
The odd integer rather than integer fillings that produces the
quantization rule (\ref{Hall-Dirac}) appears due to the
above-mentioned halved degeneracy of the $n=0$ LL. Another
interesting feature of Eq.~(\ref{sigma-steps}) (see also
Figs.~\ref{fig:1} and \ref{fig:2}) is that $\sigma_{xy} = \pm 2
e^2/h$ for the fillings $\nu_B < 1$ and it  crosses 0 only when
$\mu$ changes sign. On the contrary, in a conventional IQHE
$\sigma_{xy}=0$ for  $\nu_B < 1$.

Although Eqs.~(\ref{hall-clean}) - (\ref{sigma-steps}) are
obtained in the clean limit and using a simple bare bubble
expression for conductivity, our main result (\ref{Hall-Dirac}) is
model independent and is only based on the $n=0$ level anomaly.

Now we rewrite Eqs.~(\ref{hall-clean})-(\ref{filling}) in terms of
the Fermi distribution
\begin{equation}
\label{sigma-H-nf}
\begin{split}
\sigma_{xy}=-&\frac{2 e^2 {\rm sgn}(eB)}{h}
\sum\limits_{n=0}^\infty  (2n+1) \\
\times &\left[ n_F(M_n-\mu)+n_F(-M_n-\mu)\right.\\
& \left.  -n_F(M_{n+1}-\mu)- n_F(-M_{n+1}-\mu)\right]
\end{split}
\end{equation}
to compare it with Eq.~(18) of Ref.~\cite{Jonson1984PRB} that
was obtained for an ideal two-dimensional electron gas
\begin{equation}
\label{sigma-Jonson} \sigma_{xy} = -\frac{e^2}{h} \sum_{n=0}^\infty
(n+1)[n_F(\omega_n^{\mathrm{nonrel}}) -
n_F(\omega_{n+1}^{\mathrm{nonrel}})]
\end{equation}
with nonrelativistic spectrum
\begin{equation}
\label{nonrelLL} \omega_n^{\mathrm{nonrel}} = \frac{e \hbar B}{mc}
\left(n+ \frac{1}{2} \right).
\end{equation}

There is a commonsense reasoning
\cite{Dresselhaus2002AP,Novoselov2004Science} that graphene is  a
two-band [the first band would corresponds to the electrons with
$\omega_n = M_n-\mu$ and the second band,  to the holes with
$\omega_n = - M_n-\mu$], two-valley [corresponding to $\mathbf{K}$
and $\mathbf{K}^\prime$ points of graphen's Fermi surface]
semiconductor with zero gap $\Delta$ between the bands. Accordingly
its Hall conductivity can be directly obtained from
(\ref{sigma-Jonson}) by summing over all these bands and valleys
\begin{equation}
\label{sigma-H-semicond}
\begin{split}
\sigma_{xy}^{\mathrm{semicond}}&=-\frac{2 e^2 {\rm sgn}(eB)}{h}
\sum\limits_{n=0}^\infty  2(n+1)\\
\times & \left[n_F(M_n-\mu)+n_F(-M_n-\mu)\right. \\
& \left. -n_F(M_{n+1}-\mu)- n_F(-M_{n+1}-\mu)\right],
\end{split}
\end{equation}
where we also counted  spin degeneracy. It is easy to see that
Eqs.~(\ref{sigma-H-nf}) and (\ref{sigma-H-semicond}) correspond to
two completely different Hall conductivity quantization rules, viz.
Eq.~(\ref{sigma-H-nf}) which correctly counts the degeneracy of the
$n=0$ level produces Eq.~(\ref{Hall-Dirac}), while the
semiconducting analogy (\ref{sigma-H-semicond}) leads to
\begin{equation}
\label{Hall-semicond} \sigma_{xy}^{\mathrm{semicond}} = -
\frac{4e^2}{h}n, \qquad n =0,1,\ldots.
\end{equation}
Here we assumed that $e,B,\mu>0$. Although previous experimental
observations  supported the picture based on
Eq.~(\ref{Hall-semicond}), the latest experiments made on thin films
\cite{Geim2005} are in accord with the unconventional Hall
quantization (\ref{Hall-Dirac}). This shows that in an applied
magnetic field the semiconducting interpretation of graphene's band
structure that led us to Eq.~(\ref{sigma-H-semicond}) becomes
invalid (see also Ref.~\cite{Ando2005JPSJ}). The drastic difference
between Eqs.~(\ref{Hall-Dirac}) and (\ref{Hall-semicond}) is caused
by the above-mentioned fact that the lowest LL in Dirac theory is
special and has twice smaller degeneracy than the levels with $n>0$,
because depending on the sign $eB$ it is occupied either by
electrons or holes, while higher levels contain both electrons and
holes \cite{Gusynin1995PRD,Johnson1949PR}. In the nonrelativistic
theory when the Lande factor $g \ne 2$ all Landau levels have the
same degeneracy \cite{foot1}. It turns out that graphene for which
the valence and conduction bands intersect in discrete points
\cite{foot2}, is reasonably well described by the Dirac formalism
which naturally embodies the $n=0$ level anomaly.

We now consider the phenomenon of quantum magnetic oscillations in
graphene which is closely related to the quantization of
$\sigma_{xy}$  and discuss the specific of the $n=0$ level. The de
Haas van Alphen and Shubnikov de Haas effects in graphene were
studied in
Refs.~\cite{Sharapov2004PRB,Luk'yanchuk2004PRL,Gusynin2005PRB}. In
particular, in Ref.~\cite{Gusynin2005PRB} it was shown that the
oscillatory part of the diagonal conductivity (\ref{sigma_xx}) is
given by
\begin{equation}
\sigma_{xx} \varpropto \sum_{k=1}^{\infty} \cos \left[\frac{\pi k
\mu^2}{\hbar v_F^2 |eB|/c} \right] R_T(k) R_D(k) R_s(k),
\end{equation}
where $R_T$, $R_D$ and $R_s$ are respectively the temperature,
Dingle and spin factors. Using the relationship $\mu^2 = \pi \hbar^2
v_F^2 |\rho|$ valid for $T=\Gamma=B=0$ \cite{Gorbar2002PRB} one can
check that the minima of the diagonal conductivity (\ref{sigma_xx})
occur at the fillings $\nu_B=2n+1$ giving an indication of the
possible positions of the plateaux in the IQHE \cite{Geim2005}.
[Note that in thick films the minima of $\sigma_{xx}$ occur at
integer fillings \cite{Morozov2005}.] Obviously for $\mu=0$ there is
no oscillations of $\sigma_{xx}$, the conductivity
$\sigma_{xx}(\mu=0) = 2e^2/(\pi^2\hbar)$ becomes a field independent
universal \cite{foot3} quantity that is another distinctive feature
of the $n=0$ level anomaly.

Although the quantization (\ref{Hall-Dirac}) can be understood by
considering noninteracting Dirac quasiparticles placed in an
external magnetic field, even this simple model reveals other
unusual properties \cite{Gusynin1995PRD} intimately related to
nontrivial dynamics of  quasiparticles from the $n=0$ level. For
example, the $U(4)$ symmetry of the Lagrangian (\ref{Lagrangian}) is
spontaneously broken down to $U(2)\times U(2)$ at $\mu=0$ in
non-zero magnetic field even in the absence of additional
interaction between fermions \cite{Gusynin1995PRD} thus leading to
the emergence of the chiral condensate $\langle {\bar \Psi}
\Psi\rangle$. Including many body effects such as an attractive
interaction between quasiparticles, could further generate a gap for
quasiparticles like the above mentioned gap $\Delta$ (see e.g.
Refs.~\cite{Gorbar2002PRB,Khveshchenko2001PRL}). Fortunately in the
case of the IQHE the presence of the condensate does not affect our
consideration. On the other hand, a possible gap generation for the
fermions from the lowest LL might become important for the
fractional quantum Hall effect and this issue certainly deserves
further experimental and theoretical study.

To conclude, we have shown that the integer numbers associated
with quantized Hall conductivity in graphene have an unusual
pattern $\sigma_{xy} h/e^2 =2,6,10,14\ldots$. We argued that it is
related to the fact that a theoretical description of graphene is
based on $2+1$ dimensional Dirac theory, where the lowest Landau
level has half of the higher Landau levels degeneracy.

We are indebted to A.~Geim for showing us his experimental results
prior to publication and for stimulating discussions. We also thank
J.P.~Carbotte, V.M.~Loktev and V.A.~Miransky for useful discussions
and W.A.~de~Heer,  P.~Kim for informing us about their latest
results. S.G.Sh. was supported by the Natural Science and
Engineering Council of Canada (NSERC) and by the Canadian Institute
for Advanced Research (CIAR).


\begin{thebibliography}{99}

\bibitem{Novoselov2004Science} K.S.~Novoselov, A.K.~Geim, S.V.~Morozov,
D.~Jiang, Y.~Zhang, S.V.~Dubonos, I.V.~Grigorieva, and A.A.~Firsov,
Science {\bf 306}, 666 (2004); K.S.~Novoselov, A.K.~Geim,
S.V.~Morozov, S.V.~Dubonos, Y.~Zhang, and D.~Jiang,
cond-mat/0410631.

\bibitem{Semenoff1984PRL} G.W.~Semenoff, Phys.~Rev.~Lett. {\bf 53},
2449 (1984).

\bibitem{Kopelevich2003PRL} Y.~Kopelevich, J.H.S.~Torres, R.R.~da Silva, F.~Mrowka, H.~Kempa, and
P.~Esquinazi, Phys.~Rev.~Lett. {\bf 90}, 156402 (2003).

\bibitem{Morozov2005} S.V.~Morozov, K.S.~Novoselov, D.~Jiang, A.A.~Firsov,
S.V.~Dubonos, A.K.~Geim, cond-mat/0505319.

\bibitem{Berger2004JPCB} C.~Berger, Z.~Song, T.~Li, X.~Li, A.Y.~Ogbazghi,
R.~Feng, Z.~Dai, A.N.~Marchenkov, E.H.~Conrad, P.N.~First, and
W.A.~de Heer, J. Phys. Chem. B {\bf 108}, 19912 (2004).

\bibitem{Zhang2005PRL} Y.~Zhang, J.P.~Small, M.E.S.~Amori, and
P.~Kim,  Phys.~Rev.~Lett. {\bf 94}, 176803 (2005).

\bibitem{Sharapov2004PRB} S.G.~Sharapov, V.P.~Gusynin, and H.~Beck,
Phys.~Rev. B {\bf 69}, 075104 (2004).
%%CITATION = PHRVA,B69, 075104;%%

\bibitem{Gusynin2005PRB} V.P.~Gusynin and S.G.~Sharapov, Phys. Rev. B
{\bf 71}, 125124 (2005).
%%CITATION = COND-MAT 0411381;%%

\bibitem{Luk'yanchuk2004PRL} I.A.~Luk'yanchuk and Y.~Kopelevich,
\prl {\bf 93}, 166402 (2004).

\bibitem{Geim2005} A.K.~Geim, private communication.

\bibitem{Gusynin1995PRD} V.P.~Gusynin, V.A.~Miransky, and I.A.~Shovkovy, Phys.~Rev.~Lett.
{\bf 73}, 3499 (1994);
%%CITATION = HEP-PH 9405262;%%
Phys.~Rev.~D {\bf 52}, 4718 (1995).
%%CITATION = HEP-TH 9407168;%%

\bibitem{foot1} Note that in the relativistic $(3+1)$
Dirac theory the Zeeman term is built in the formalism. When the
relativistic eigenenergy, $E(n,\sigma=\pm,k_z=0) = [ m^2c^4 + mc^2
\hbar |eB|/(mc) (2n+1+\sigma) ]^{1/2}$ is separated into the Pauli
term with $\sigma \mu_B B$ and the diamagnetic term, the
nonrelativistic LL with the same degeneracy and the spectrum given
by Eq.~(\ref{nonrelLL}) emerge. Each nonrelativisitic LL suffers a
Zeeman splitting $\pm 1/2 g \mu_B B$, so that it doubles the number
of LL. Here we included the Lande factor $g$ which plays an
important role in solids. For $g=2$, there is an accidental (from a
nonrelativistic point of view) degeneracy of LL, so that each level
with $n>0$ is filled by both spin down and spin up electrons from
the $n-1$ level. The $n=0$ LL is special and remains nondegenerate,
so that we come to the picture considered in this paper. In our
case, however, an ``accidental'' degeneracy of $n>0$ LL is not
related to the spin degree of freedom and occurs due to the presence
of two sublattices in graphene.


\bibitem{Gorbar2002PRB} E.V.~Gorbar, V.P.~Gusynin, V.A.~Miransky,
and I.A.~Shovkovy, Phys. Rev. B {\bf 66}, 045108 (2002).
%%CITATION = COND-MAT 0202422;%%

\bibitem{Gusynin2005} The corresponding analytical expression
for $\mathcal{A}_{xy}(\omega,B,\Gamma)$ is simple, but rather
lengthy, V.P.~Gusynin and S.G.~Sharapov, in preparation.

\bibitem{Zheng2002PRB} Similar figures also appear in
Y.~Zheng and T.~Ando, Phys. Rev. B {\bf 65}, 245420 (2002), but with
a twice smaller size of the steps of $\sigma_{xy}$, and without
making a link between the unusual behavior of $\sigma_{xy}$ and the
quantum anomaly of the $n=0$ level.

\bibitem{Schakel1991PRD} A.M.J.~Schakel, Phys.~Rev.~D {\bf 43}, 1428
(1991).

\bibitem{Khveshchenko2001PRL} D.V.~Khveshchenko,
Phys.~Rev.~Lett. {\bf 87}, 206401; {\em ibid.} {\bf 87}, 246802
(2001).

\bibitem{Hajdu.book}  M.~Jan$\beta$en, O.~Veihweger, U.~Fastenrath, and J.~Hajdu,
Introduction to the Theory of the Integer Quantum Hall Effect,
edited by J.~Hajdu, VCH, Weinheim, 1994.

\bibitem{Jonson1984PRB} M.~Jonson and S.~M.~Girvin, Phys. Rev. B {\bf 29}, 1939 (1984).

\bibitem{Dresselhaus2002AP} M.S. Dresselhaus and G. Dresselhaus,
Adv.~Phys. {\bf 51}, 1 (2002).

\bibitem{Ando2005JPSJ} T.~Ando, J.~Phys.~Soc.~Jpn. {\bf 74}, 777
(2005).

\bibitem{Johnson1949PR} M.H.~Johnson and B.A.~Lippmann, Phys.~Rev.
{\bf 76}, 828 (1949).

\bibitem{foot2} The points near
which the electrons have a linear spectrum and are described by a
two-level Hamiltonian are called in the literature
\cite{Schakel1991PRD}  {\em diabolic points \/}. There is a theorem
of H.B.~Nielsen and M.~Ninomiya, Nucl.~Phys.~{\bf B185}, 20 (1981);
{\bf B 193}, 173 (1981) that asserts that in condensed matter
systems diabolic points come in parity-inavariant pairs. Accordingly
the Dirac Lagrangian (\ref{Lagrangian}) which embedes a pair of such
points, written using a parity preserving $4\times4$ reducible
representation of $\gamma$ matrices is rather natural.

\bibitem{foot3} This universality has the same origin as the
universality considered in E.~Fradkin, Phys.~Rev.~B {\bf 33}, 3263
(1986) for degenerate semiconductors and in P.A.~Lee, Phys. Rev.
Lett. {\bf 71}, 1887 (1993) for $d$-wave superconductors. Moreover
in the case of graphene where the quasiparticles have not only
linear disperion relation similar to the Bogolyubov quasiparticles
in $d$-wave superconductor, but also couple to the vector
potential in the conventional QED$_{2+1}$ way, the conductivity
$\sigma_{xx}(\mu=0)$ is expected to be field independent
\cite{Gorbar2002PRB}.


\end{thebibliography}
\end{document}